\documentclass{svproc}
\usepackage{graphicx}
\usepackage{url,float}

\begin{document}
\mainmatter              
\title{Identifying  Discrete Breathers Using Convolutional Neural Networks}
\titlerunning{Discrete Breathers with CNNs}  
%
\author{T.  Dogkas,  M.  Eleftheriou,  G.  D.  Barmparis, and G.  P.  Tsironis$*$}
\authorrunning{T. Dogkas et al.} 
%
%
\institute{Department of Physics, University of Crete, P. O. Box 2208, Heraklion 71003, Greece\\
\email{$*$gts@physics.uoc.gr}}

\maketitle

\begin{abstract}
Artificial intelligence in the form of  deep learning is now very popular and directly implemented in many areas of science and technology.  In the present  work we study time evolution of Discrete Breathers  in one-dimensional nonlinear chains using the framework of Convolutional Neural Networks.  We focus on differentiating discrete breathers which are localized nonlinear modes  from linearized phonon modes. The breathers are localized in space and time-periodic solutions of non-linear discrete lattices while phonons are the linear collective oscillations of interacting atoms and molecules.  We show that deep learning neural networks  are indeed able not only to distinguish breather from phonon modes but also determine with high accuracy the underlying nonlinear on-site potentials that generate breathers.  This work can have extensions to more complex natural systems.

\end{abstract}

\parindent=0pt
\parskip=0.5em

\def\thesection{\arabic{section}}
\section{Introduction}\label{sec:1}

Discrete breathers (DBs) or Intrinsic Localized Modes (ILMs) are time periodic and space localized modes that appear in discrete nonlinear lattices \cite{ST}.  During the last over thirty year period there has been substantial amount of theoretical and experimental work that generated a body of precise knowledge regarding these modes \cite{FW}. In the present work we use modern tools of Machine Learning (ML) in order to address an entirely different question, viz. whether DBs can be recognized in some automatic form without the need of direct human intervention.  We believe that this is a significant question since its precise answer may lead to much easier and direct detection of nonlinear modes in natural systems.   In this preliminary work we sharpen the question to: Is it possible to recognize DBs and phonon modes in simple one-dimensional chains with nonlinear on-site potentials using Convolutional Neural Networks (CNNs)? In order to construct DBs we use the numerically exact method introduced by Aubry for generation from the anticontinuous limit \cite{Aubry}.

In this brief account we have four sections.  In the first section we give some details on the generation of DBs and phonons samples to be used subsequently.  In the second section a CNN model is developed in order to distinguish DBs from phonons.  In the following section a CNN model is developed to identify the on-site nonlinear potential through which the linear and nonlinear modes were generated.  Finally,  in the last section we conclude and give some further perspectives of this work.

\section{Creation of breather and phonon samples}

For the ML analysis in this work we create 459 samples of breathers and phonons using the anticontinuous limit method in 1D lattice with the Hamiltonian \cite{ME,LET,LT} and three different nonlinear on-site potentials as outlined in Table \ref{tab : 1}. The DBs have frequencies outside the phonon spectrum and their stability is checked through Floquet analysis. We give some details on the methods used below.

\begin{table}
\begin{center}
\begin{tabular}{ |p{3cm}|p{5cm}|}
\hline
\multicolumn{2}{|c|}{\textbf{Potentials}}\\
\hline
\textbf{Hard $\phi^4$} &$ V(x) = \frac{x^{2}}{2}+\frac{x^{4}}{4}$ \\
\textbf{Morse} & $ V(x) = - \frac{1}{2}(1-e^{-x})^2$\\
\textbf{Double Well}  & $ V(x) = - \frac{(x-1)^{2}}{2}+\frac{(x-1)^{4}}{4}$\\
\hline
\end{tabular}
\end{center}
\caption{\label{tab : 1} The three nonlinear on-site potentials used in this work,  are the hard $\phi^4$,  the Morse and double-well potentials.}
\end{table}

\medskip\textbf{The breather solution}

We outline very briefly the procedure we follow in order to obtain numerically exact breathers and make sure they are stable.  We use a  Hamiltonian in the form:
\begin{equation}
H= \sum_{N}\frac{p_{n}}{2} + V(x_{n})+W(x_{n})
\label{EQ-1}
\end{equation}
where $x_{n}$ the displacement at the node $n$ of the 1D lattice,  $ p_{n}$  is the corresponding momentum,  $V$ is one of the three nonlinear on-site potentials used, while $W$ is the interaction potential, viz.  $W(x_{n})= k (x_{n-1}+x_{n+1}-2x_{n})$.  The value of the parameter $k$ is quite important since it affects the existence as well as the shape of DBs. 

In the numerical procedure we first obtain the breather solution (see below) and subsequently we linearize the equations of motion around this solution.  The small amplitude plane waves of the form $ x_{n} (t) = e^{i(\omega_{b}t-qn)}$ propagate with frequencies:

\begin{equation}
\omega_{b} = V^{''}(0)+4W^{''}(0)\sin (\frac{q}{n})
\label{EQ-2}
\end{equation}

\begin{figure}
\includegraphics[width=\textwidth]{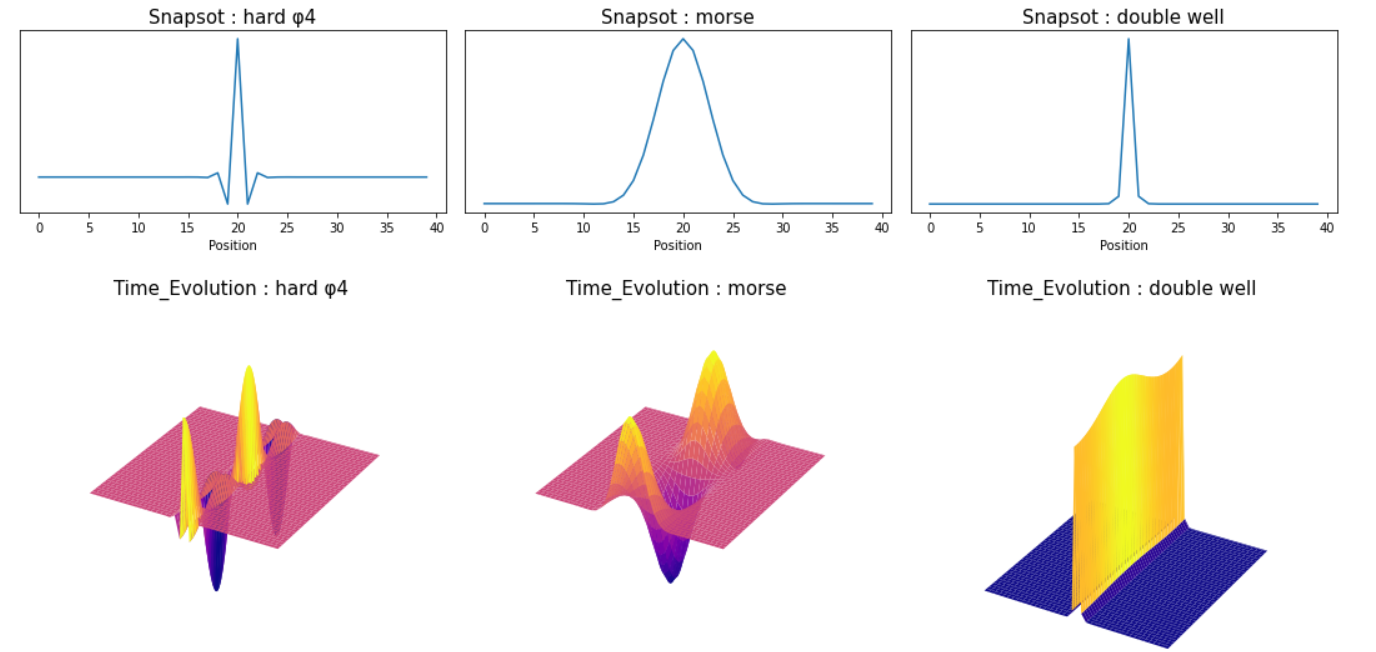}
\caption{\label{fig : breathers }Breather images and their time evolution for hard $\phi^4$,  Morse and double well potentials with coupling $k=0.1$ and frequencies $f=1.317,$ $f=0.967$ and $f=0.949$ respectively.}
\end{figure}

We solve the full equations of motion at zero coupling in order obtain a trivial breather with specific amplitude-period relationship \cite{Aubry,ME}. Subsequently,  we create a vector $ U(x_{1},x_{2},\cdots,x_{n},p_{1},p_{2},p_{n})$ for the N lattice sites (we take $N=40$)  and set in the middle of the lattice the initial condition $x_{n}=0.9$ and $p_{n}=0$, i.e the trivial breather.  Subsequently, we  increase the coupling $k$ in small steps and solve the equations of motion; this iterative procedure is repeated until the desired coupling value is reached.  Since the DBs are time periodic solutions,  for periodic evolution derived from the map $T$ we have $U_{k} + \Delta = T(U_{k}+\Delta) \Leftrightarrow U_{k} + \Delta = T(U_{k}) +\partial{T} x \Delta$, where $\partial{T}=M$ the tangent map of the system.  By minimizing the square of norm $||T(U) + M\Delta -(U+\Delta) || ^ {2} $  we obtain the matrix $\Delta$ that gives the final solution for the breather and its stability.  We repeat this procedure in with each breather sample and thus obtain stable DB solutions as shown in Fig. \ref{fig : breathers }.

\bigskip\textbf{Phonon spectrum}

\begin{figure}
\begin{center}
\includegraphics[width=0.7\textwidth]{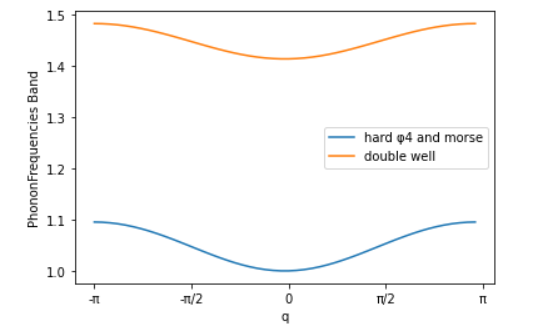}
\caption{\label{fig : spectrum} Phonon Frequency band with upper limit $\omega_{b} = 1.095$, lower limit $\omega_{b'} = 1$ and coupling $k=0.05$ for hard $\phi^4$ and Morse potentials and with upper limit $\omega_{b}= 1.483$, lower limit $\omega_{b'}= 1.414$.}
\end{center}
\end{figure}

\begin{figure}
\includegraphics[width=\textwidth]{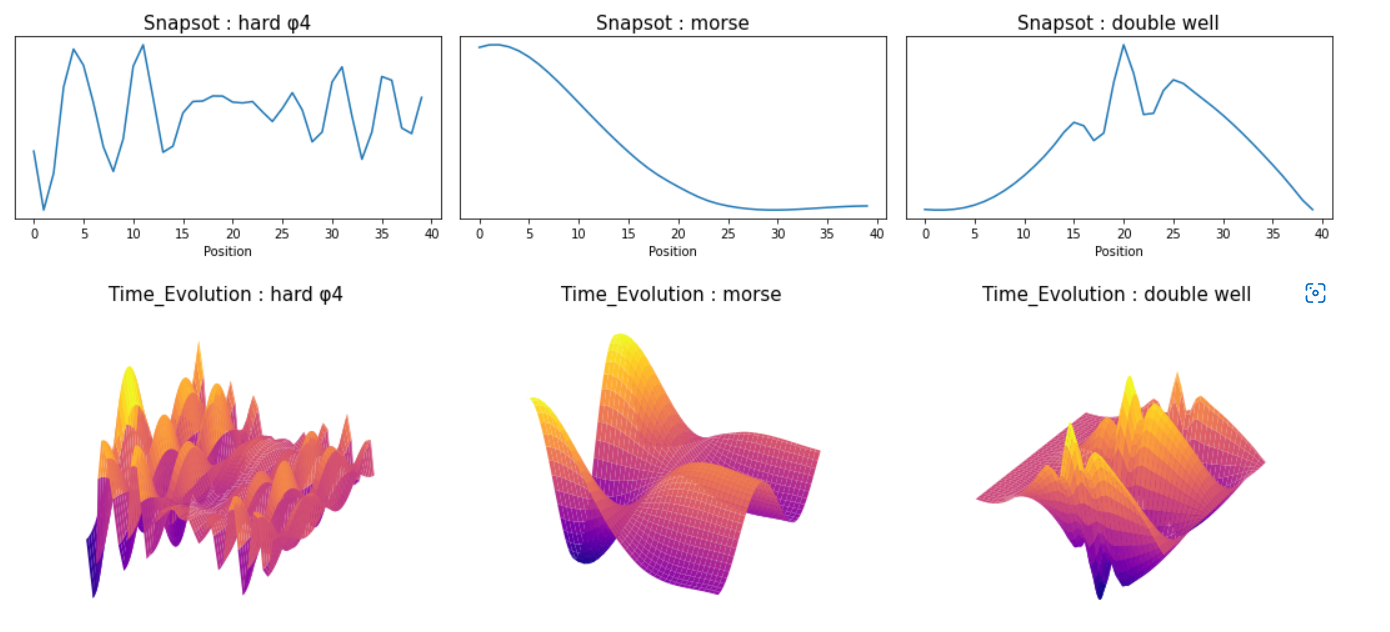}
\caption{\label{fig : phonons} Phonon images and their time evolution for potentials hard $\phi^4$,  Morse and double well with coupling $k=0.1$ and frequencies $f=1.042$, $f=1.001$, $f=1.415$ respectively.}
\end{figure}

The phonon spectrum for different couplings is given by equation \ref{EQ-2}.  We designate the upper limit of the phonon band as $\omega_b$, while the lower limit as $\omega_b'$.  We get acoustic-like breathers when the breather frequency, $\Omega_{b}$, is less than $\omega_{b'}$ and optic-like breathers when $\Omega_{b} > \omega_{b}$.  In Fig. \ref{fig : spectrum}, we present the phonon frequency band with upper limit $\omega_{b} = 1.095$,  lower limit $\omega_{b'} = 1$ and coupling $k=0.05$ for the hard $\phi^4$, and with upper limit $\omega_{b}= 1.483$, lower limit $\omega_{b'}= 1.414$ for the Morse potential, respectively.  Snapshots and the time evolution of phonons for the hard $\phi^4$, Morse and double well potential with coupling $k=0.1$, and frequencies $f=1.042$, $f=1.001$, and $f=1.415$, respectively, are shown in Fig. \ref{fig : phonons}.

\newpage

\bigskip\textbf{Stability of Discrete Breathers}

\begin{figure}
\begin{center}
\includegraphics[width=0.6\textwidth]{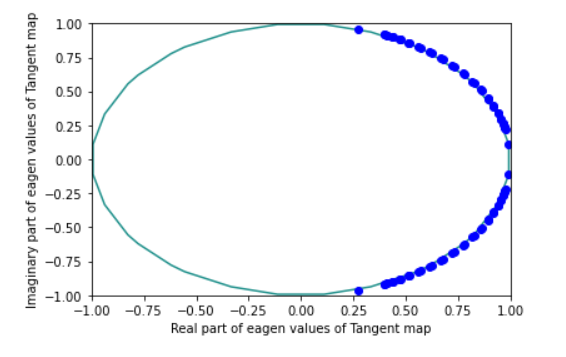}
\caption{\label{fig : Floquet}  Eigenvalues of tangent map M matrix of a breather with potential hard $\phi^4$, frequency $f=1.227$ and coupling $k=0.1.$}
\end{center}
\end{figure}

\begin{figure}
\begin{center}
\includegraphics[width=0.8\textwidth]{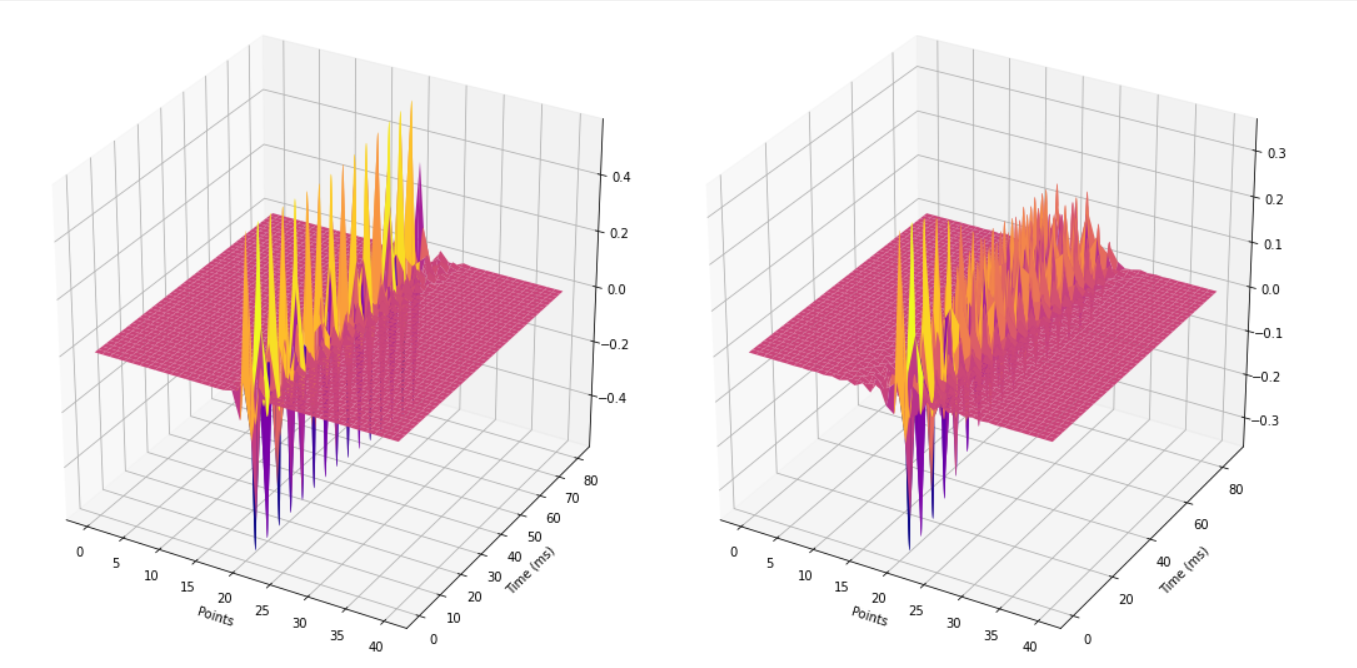}
\caption{\label{fig : SU} Images of the time evolution over 15 period of time with coupling $k=0.05$ and hard $\phi^4$ potential of a stable breather with frequency $ f = 1.099$ (left) and an unstable breather with frequency $f = 1.17$ (right).}
\end{center}
\end{figure}

We use two different methods to examine the stability of DBs. The first method is the Floquet analysis (Fig. \ref{fig : Floquet}), where we obtain the eigenvalues of the tangent map M.  Stable DBs solutions give Floquet eigenvalues where their imaginary and real part lie on a circle with radius r=1.  The second method is by direct observation the time evolution of DBs. In Fig. \ref{fig : SU}, we present the time evolution of a stable (left) and an unstable (right) breather for the hard $\phi^4$ potential.

\section{Machine Learning Training Process and Results}

\begin{figure}
\includegraphics[width=\textwidth]{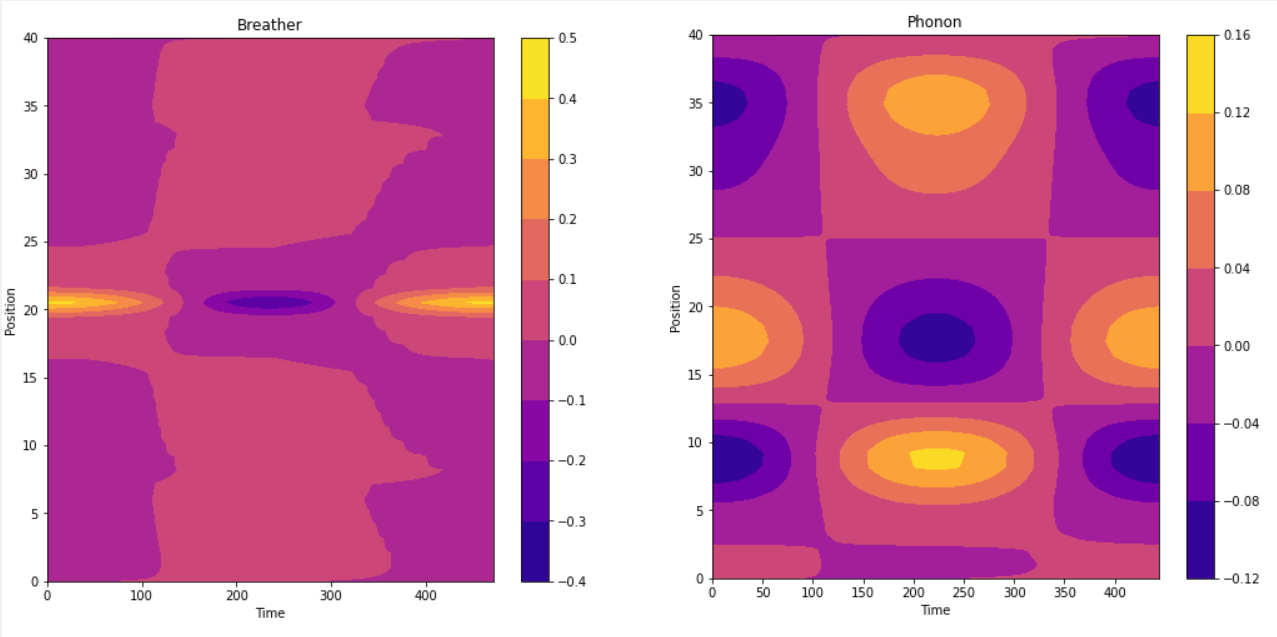}
\caption{\label{fig:BP_Contour} Contour plots of the time evolution of a breather with frequency $f=1.332$ and a phonon with frequency $f=1.4161$, with coupling $k=0.1$ for the double well potential, respectively. The x-axis represents the time,  the y-axis the position of each node of the 1D chain and the colorbar the amplitude.}
\end{figure}

We create a dataset of 459 samples,  with equally distributed phonons and DBs.  Each sample is a single channel gray-scaled, 2D image of size 40x822 pixels, of the time evolution of a DB or phonons.  Each image was constructed in a way that at least one time period is included.  In Fig. \ref{fig:BP_Contour} we present a colorized sample of each category of DBs and phonons. The dataset was shuffled to avoid biases. Twenty percent of the samples were separated to create a hold-out set for testing. The rest 80\% of the samples was split to create a training (80\%) and a validation (20\%) set.  The dataset was normalized using the ImageDataGenerator package \cite{HTF}.  Two models of identical CNNs were created.  The features extractor part of each model consisted of 3 convolutional layers with 32, 64 and 64 (3x3) kernels, respectively and a \textit{relu} activation function.  The first two convolutional layers in each model were followed by a (2x2) Max-pooling layer.  The classifier of each model contains two fully connected layers with 64 and 2 nodes respectively, for the case of the DB/phonons classifier,  and 64 and 3 nodes layers for the three potentials classification.  A \textit{relu} activation function was used for the first layer and a \textit{softmax} one for the output layer of each classifier. The models were allowed to train for 100 epochs, while an early stopping criterion with patience 5 epochs was monitoring the validation error in order to avoid over-fitting.  The f1-score was used to evaluate the performance of the models.


\newpage

\begin{figure}
\includegraphics[width=\textwidth]{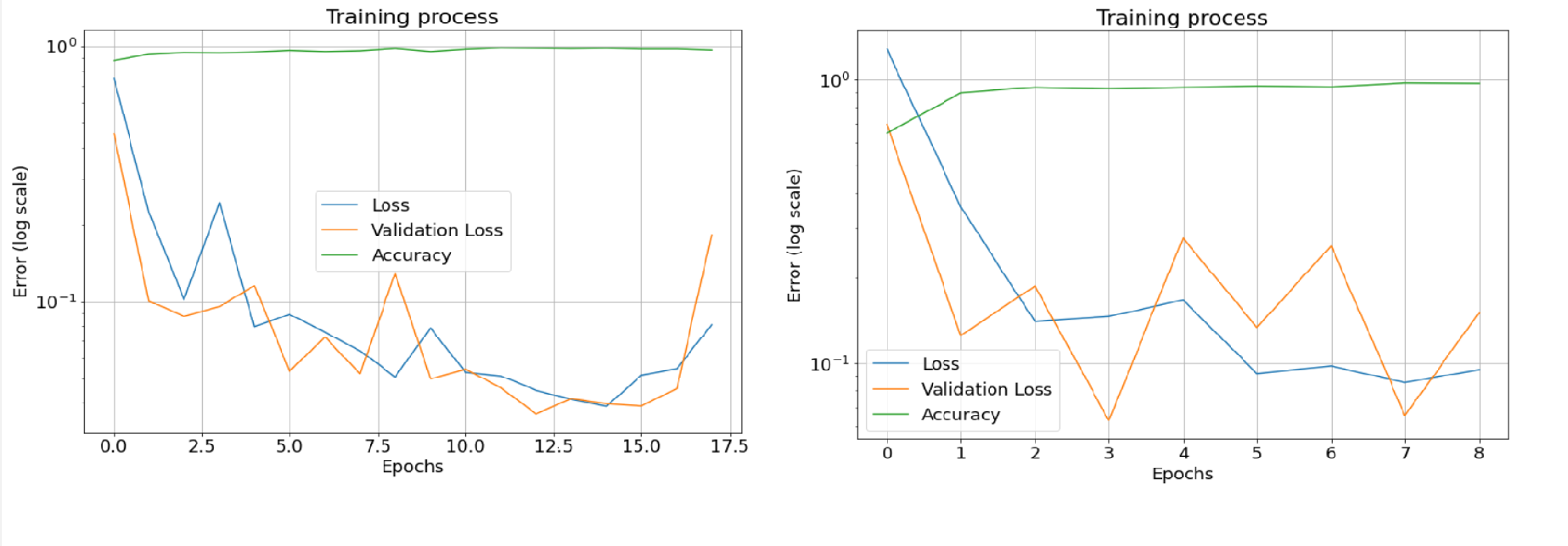}
\caption{\label{fig:Training1} The training accuracy and loss and the validation loss as a function of the number of epochs, for the breather-phonon classification (left) and the potential classification (right) model respectively.}
\end{figure}

\begin{figure}
\includegraphics[width=0.9\textwidth]{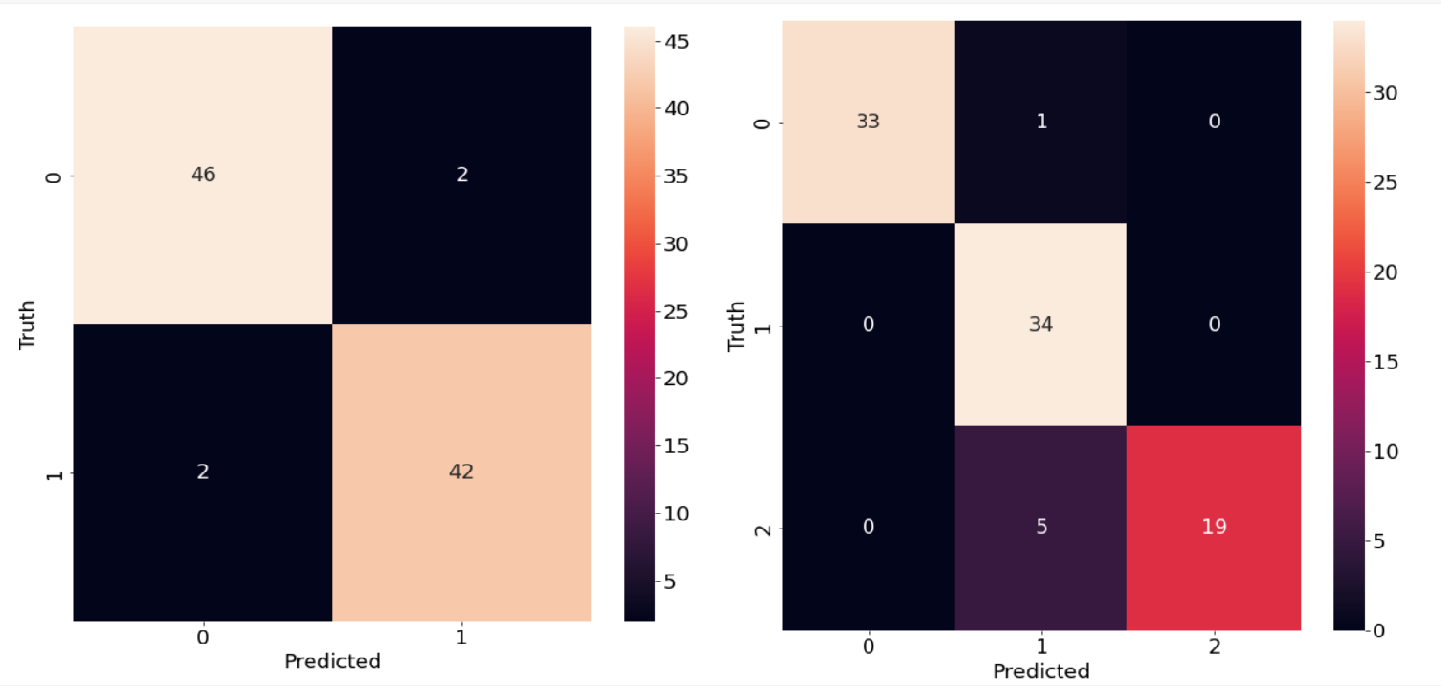}
\caption{\label{fig:Training2} The confusion matrix of the hold-out test set for the breather-phonon classification (left) and the potential classification (right) model respectively.  In the left plot '0' indicates phonons, while '1' indicates breathers.  In the right plot '0' indicates Morse, '1' indicates double well, and '2' indicates hard $\phi^4$ potential, respectively. }
\end{figure}

In Figs. \ref{fig:Training1}, \ref{fig:Training2} and in Table \ref{tab : 2},  we  present the training process and the results of our analysis.  Both the $f1$ score is high while the confusion matrix shows very high degree of classification accuracy. 

\begin{table}
\begin{center}
\begin{tabular}{ |p{3cm}|p{3cm}|p{3cm}|}
\hline
\multicolumn{3}{|c|}{\textbf{Test set - Classification Results}}\\
\hline
 &\textbf{Breather-Phonon}& \textbf{Potentials} \\
\hline
\textbf{Accuracy} & 95.7\% & 93.5\% \\
\textbf{f1-score} & $0.955$ & $0.934$\\
\hline
\end{tabular}
\end{center}
\caption{\label{tab : 2}Table of the accuracy and the f1-score of the test set for each model.}
\end{table}

\newpage

\section{Conclusions}

This preliminary work that applies CNNs to the breather-phonon classification problem gives promising results as a previous ML work did in chaotic systems\cite{ABT}.  We note that  similar results were also found through the use of different ML methods in the breather problem \cite{BK}.  We find here that it is indeed possible to perform an accurate classification based on the assumptions presented previously.  The quantitative result of the confusion matrix on the test set shows that the classification is excellent provided the CNNs are trained appropriately. This result, however, is not unexpected since breathers and phonons as 2D images can be easily distinguished by a human eye.  What is more noteworthy,  however,  is the possibility to  find the underlying potential these modes stem from.   Indeed,  we see that with proper training the ML model may distinguish the specifics of the underlying dynamics.  This strong feature opens up the possibility for a deeper use of Deep Learning in nonlinear physics.   It is not unreasonable to expect that under specific,  controlled conditions,  we may be able to use experimental data to infer the precise underlying dynamics of a complex system.   We note that similar indications are also found in the study of chaotic systems with ML methods \cite{ABT}.  We conclude that a more detailed study of the potential of machine learning in complex systems is necessary.



\end{document}